\documentclass[12pt,a4paper,lineno]{article}
\usepackage[T1]{fontenc}
\usepackage{physics}
\usepackage[utf8]{inputenc}
\usepackage[affil-it]{authblk} 
\usepackage{etoolbox}
\usepackage{lmodern}
\usepackage{xcolor}
\usepackage{graphicx}
\usepackage{titletoc}
\usepackage{xr}

\usepackage{float}
\usepackage[font=small,labelfont=bf]{caption}
\usepackage{hyperref}
\hypersetup{
    colorlinks=true,
    linkcolor=blue,
    citecolor=blue,
    filecolor=blue,      
    urlcolor=blue,
    pdftitle={Simultaneous High-Fidelity Single-Qubit Gates in a Spin~Qubit~Array},
    pdfauthor={Leon C. Camenzind},
    pdfpagemode=UseNone,
    }

\usepackage{geometry}
\usepackage{pdfpages} 
\usepackage{pgffor} 

\makeatletter
\AtBeginDocument{\let\LS@rot\@undefined}
\makeatother

\usepackage{lineno}
\usepackage{etoolbox}

\newif\ifarXiv
\arXivtrue 

\title{\large\textbf{Simultaneous High-Fidelity Single-Qubit Gates in a Spin~Qubit~Array} }

\author[*,1,2,${\dagger}$]{Yi-Hsien Wu}
\author[*,1,${\dagger}$]{Leon C. Camenzind}
\author[1,3]{Patrick B\"utler}
\author[1]{Ik Kyeong Jin}
\author[1]{Akito Noiri}
\author[1]{Kenta Takeda}
\author[1]{Takashi Nakajima}
\author[4]{Takashi Kobayashi}
\author[5]{Giordano Scappucci}
\author[2,6,7]{Hsi-Sheng~Goan}
\author[1,4,${\dagger}$]{Seigo Tarucha}

\affil[1]{Center for Emergent Matter Science, RIKEN, 2-1 Hirosawa, Wako-shi, 351-0198, Saitama, Japan}
\affil[2]{Department of Physics, National Taiwan University, Taipei 10617, Taiwan}
\affil[3]{Laboratory for Solid State Physics, ETH Zurich, CH-8093 Zurich, Switzerland}
\affil[4]{Center for Quantum Computing, RIKEN, 2-1 Hirosawa, Wako-shi, 351-0198, Saitama, Japan}
\affil[5]{QuTech and Kavli Institute of Nanoscience, Delft University of Technology, 2600 GA Delft, The Netherlands}
\affil[6]{Center of Quantum Science and Engineering, National Taiwan University, Taipei, 10617, Taiwan}
\affil[7]{Physics Division, National Center for Theoretical Sciences, Taipei, 10617, Taiwan}
\affil[*]{These authors contributed equally to this work.}
\affil[$\dagger$]{Correspondence to L.C.C.\ (\texttt{leon.camenzind@riken.jp}), \newline Y.-H.W.~(\texttt{yi-hsien.wu@a.riken.jp}), or~S.T.~(\texttt{tarucha@riken.jp}).}
\date{\small (Dated: \today)} 

\begin{document}
\maketitle
\vspace{-1.5cm}


\begin{abstract}
Silicon spin qubits offer a promising path to scalable quantum computing due to their compatibility with industrial semiconductor manufacturing and recent advances in multi-qubit integration. A key requirement for scaling quantum processors is the ability to perform high-fidelity operations in parallel across many qubits. In silicon spin systems, however, simultaneous control remains a major challenge, as fidelities typically degrade under parallel operation. In a five-qubit silicon spin array, we identify microwave-drive-induced AC Stark shifts as the dominant source of this degradation. We address this by introducing a scalable mitigation protocol based solely on pairwise phase calibrations. Using tailored control pulses on a shared control line, we achieve primitive $\pi/2$ gate fidelities well above 99.99\% for each qubit individually, with some approaching 99.999\%, surpassing previously reported fidelities in silicon spin qubits. Crucially, these fidelities are preserved above 99.99\% during simultaneous operation of up to three qubits. During parallel five-qubit operation, fidelities remain at the practical fault-tolerant threshold of 99.9\%, with the loss attributed to drive-induced decoherence resulting from increased microwave power. This effect can be mitigated through device-level improvements. By demonstrating that high-fidelity control is maintained during simultaneous operation, we overcome a central challenge in silicon spin qubits and highlight the potential of shared qubit-control lines for scaling.

\end{abstract}
\section{Main}
Quantum computing holds the potential to revolutionize fields from cryptography to materials discovery, but building scalable, fault-tolerant quantum processors remains a central challenge. Spin qubits in solid-state devices \cite{loss_quantum_1998, burkard_semiconductor_2023} are a leading platform candidate, with a compact footprint and compatibility with industrial semiconductor fabrication \cite{neyens_probing_2024, elsayed_low_2024, zwerver_qubits_2022, camenzind_spin_2022}, offering a route toward scaling to the millions of qubits envisioned for fault-tolerant quantum computing architectures \cite{Gidney2021_factorRSA}.

For fault-tolerant quantum computing, gate fidelities must exceed the threshold set by quantum error correction protocols, such as the surface code \cite{fowler_surface_2012}. While the requirement depends on the noise model, fidelities above 99.9\% are widely targeted for practical fault-tolerant operation \cite{wang_surface_2011, Google2021_ExpSuppression}. In scalable quantum processors, control must extend beyond sequential single-qubit gates to the simultaneous operation of multiple qubits, since idle qubits dephase much faster than driven ones. In our silicon spin qubits, for example, driven coherence times are about $T_2^R \sim 100~\mu$s, while idle dephasing times are only $T_{2}^{*} \sim 10~\mu$s.  This makes sequential operations impractical at scale since minimizing qubit idle times requires simultaneous control across many qubits. However, parallel qubit control typically degrades fidelities because of crosstalk. Such degradation has not only been observed in spin qubits \cite{xue_benchmarking_2019, mills_two-qubit_2022, lawrie_simultaneous_2023} but also in superconducting systems, where mitigation strategies have been developed \cite{Gambetta2012, McKay2019, Mundada2019, Zhao2022}. In spin qubits, the problem is particularly severe: fidelities were recently reported to drop considerably from 99.95\% to 99.5\% under parallel operation of only two qubits in silicon/silicon-germanium devices, the same platform used in this work \cite{mills_two-qubit_2022}. The physical origin of this degradation has remained unclear \cite{xue_benchmarking_2019, mills_two-qubit_2022, lawrie_simultaneous_2023}. Addressing it is therefore essential for implementing large-scale error correction codes that require parallel gate execution.

We identify microwave-drive-induced AC Stark shifts as the dominant source of fidelity degradation during parallel operation and introduce a compact calibration protocol that compensates these shifts using only pairwise measurements. This approach avoids the exponential overhead of full crosstalk characterization and enables high-fidelity parallel control of five qubits through a shared qubit-control line. Tailored control pulses allow individual qubits to reach fidelities well above 99.99\%, with some approaching 99.999\%, as measured by randomized benchmarking. Fidelities above 99.99\% are also sustained during simultaneous operation of up to three qubits, while full five-qubit parallel operation maintains fidelities at the practical fault-tolerant threshold of 99.9\% or higher. This performance exceeds previously reported fidelities in silicon spin qubits and sets a new benchmark for scalable, high-fidelity parallel control in semiconductor quantum processors. The use of a shared control line further reduces wiring complexity, supporting continued scaling toward larger qubit arrays in silicon.

\subsection{Five-qubit device and operation}\label{result:device}
\begin{figure}[H]
\centering
\captionsetup{font=footnotesize,skip=0pt,width=1\linewidth}
\includegraphics[width=1\textwidth]{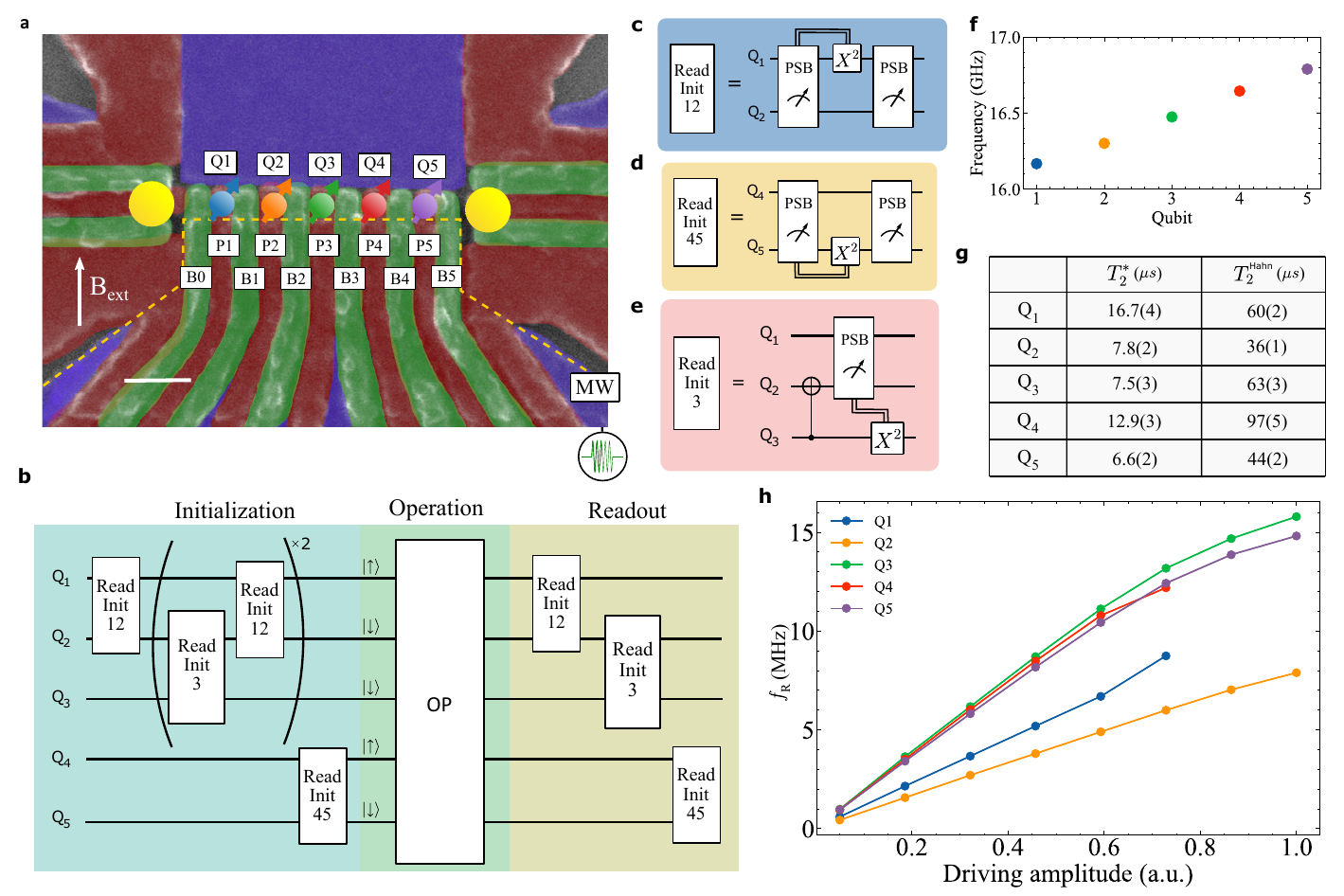}
\caption{\textbf{Five-qubit device.} 
\textbf{a} False colored scanning electron micrograph of a co-fabricated device identical to the one used in this experiment (scale bar 100 nm). An array of five quantum dots is accumulated beneath plunger gates $P_{1\sim 5}$, hosting the five spin qubits. Two sensor quantum dots (yellow) are accumulated on both ends of the array, allowing for single-shot readout. Microwave signals are applied to the MW screening gate (outlined by the dashed line) to perform single-qubit operations via electric dipole spin resonance (EDSR). A cobalt micromagnet is fabricated on top of the device (not shown) to enable qubit addressability and EDSR operations within a gradient field.
\textbf{b} Device operation scheme. The five qubits are initialized in the $|\uparrow\downarrow\downarrow\uparrow\downarrow\rangle$ state, manipulated during the operation stage using control pulses, and then read out.
\textbf{c, d} Initialization and measurement of qubit pairs Q$_{1,2}$ and Q$_{4,5}$ using PSB readout and a feedback X$^{2}$ gate conditioned on the PSB readout result. 
\textbf{e} Initialization and measurement of qubit Q$_{3}$ using QND readout and a conditional X$^{2}$ gate. 
\textbf{f} Qubit resonance frequencies at an external field of $B=0.42$ T. The micromagnet gradient induces an approximate 150 MHz separation between all adjacent qubits, enabling individual qubit addressability with MW pulses.
\textbf{g} Qubit phase coherence time $T_{2}^{*}$ and Hahn echo sequence coherence time $T_{2}^{\text{Hahn}}$ for all qubits. The total integration time is a few minutes for all measurements.
\textbf{h} The driving amplitude dependence of the qubit Rabi frequencies $f_{\text{R}}$ shows a linear relationship for all qubits up to $f_{\text{R}} \sim 8$ MHz, after which some qubits begin to exhibit nonlinear behavior.
}\label{fig1}
\end{figure}

Fig.~\ref{fig1} (a) shows the false-colored scanning electron micrograph of a co-fabricated silicon five-qubit device identical to the one used in this experiment. The device is fabricated in an isotopically purified {$^{28}$Si}/SiGe quantum well. Three overlapping aluminum metal gate layers, separated by native oxide, are deposited on the wafer to define and control an array of five quantum dots that host spin qubits. Two adjacent charge sensor quantum dots at the ends of the array are used for qubit readout \cite{philips_universal_2022}. A cobalt micromagnet (not shown) is deposited on top of the device to induce a gradient field for qubit addressability and operation \cite{kawakami_electrical_2014, yoneda_quantum-dot_2018}. A microwave signal (MW) is applied to the MW gate electrode (dashed line) to manipulate the spin states of all five qubits in the array via electric dipole spin resonance (EDSR) \cite{Golovach_2006_EDSR, Tokura2006_slantingfield, Pioro-Ladriere_2008}, thereby providing a shared control line for all qubits. Further details on the device are given in Methods.

Fig.~\ref{fig1} (b) shows the device operation scheme. We operate the device in the (3,1,1,3,1)-electron charge configuration for the five quantum dots. Pauli-spin blockade (PSB) is used to read out the parity of qubit pairs Q$_{1,2}$ and Q$_{4,5}$ by projecting qubit Q$_{2}$ (Q$_{5}$) onto qubit Q$_{1}$ (Q$_{4}$). The dots containing three electrons serve as the projection dots for the PSB readouts, which helps increase the readout windows \cite{philips_universal_2022}. The PSB feedback initialization circuits for qubit pairs Q$_{1,2}$ and Q$_{4,5}$ are shown in Fig.~\ref{fig1} (c) and (d). The qubit parity is first read out using PSB. If an even parity is measured, a conditional X$^{2}$ gate applies a $\pi$ flip to the qubits, bringing them into the $|\uparrow\downarrow\rangle$ or $|\downarrow\uparrow\rangle$ state. A second PSB readout then projects these states into the singlet $S(4,0)$ state. If an odd parity is measured, the conditional X$^{2}$ gate is not applied, and the qubit pair remains in $S(4,0)$ after the first PSB readout. The micromagnet gradient lifts the degeneracy between the spin states, allowing initialization into the $|\uparrow\downarrow\rangle$ state via an adiabatic ramp. 
Qubit Q$_{3}$ is measured by quantum non-demolition (QND) measurement between Q$_{2}$ and Q$_{3}$ \cite{yoneda_quantum_non-demolition_2020}, as shown in Fig.~\ref{fig1} (e). A CROT gate is applied with qubit Q$_{3}$ as the control qubit and Q$_{2}$ as the target. 
A PSB readout of the Q$_{1,2}$ pair is then performed, and if an even (odd) parity state is measured, we infer that qubit Q$_{3}$ is in the $|\uparrow\rangle$ ($|\downarrow\rangle$) state. Based on this readout, a conditional X$^{2}$ gate is applied to qubit Q$_{3}$, preparing it in the $|\downarrow\rangle$ state. After the QND readout, the Q$_{1,2}$ pair is re-initialized using the PSB feedback initialization circuit. The QND readout is repeated twice to increase the initialization fidelity of qubit Q$_{3}$.

The five-qubit state is initialized to $\ket{\uparrow\downarrow\downarrow\uparrow\downarrow}$ before the operation stage, during which single- and two-qubit operations are applied to control the quantum processor. Afterward, the parity of qubit pairs Q$_{1,2}$  and Q$_{4,5}$, along with the state of Q$_{3}$, is read out, yielding three bits of information per cycle. 
The resonance frequencies of the five qubits, shown in Fig.~\ref{fig1} (f), exhibit a Zeeman energy difference of $\Delta E_{\text{Z}} \sim 150$~MHz between neighboring qubits, consistent with our micromagnet design. {Coherence properties are characterized by an average dephasing time $T_{2}^{*}\sim 10.3$~$\mu\text{s}$ and an average Hahn echo decay time $T_{2}^{\text{Hahn}}\sim 60$~$\mu\text{s}$, as shown in Fig.~\ref{fig1} (g).} 
{Low residual exchange is crucial for high-fidelity single-qubit gates in multi-qubit systems \cite{rimbach-russ_simple_2023, noiri_a-shuttling_2022}. Under single-qubit operating conditions, we measure the residual exchange below $J < 16.6$~kHz for all neighboring qubit pairs (Extended Data Fig.~\ref{extend:res_exchange}).} For two-qubit gate operations, the exchange coupling is {activated} into the MHz range by pulsing the voltages on the interdot barrier gates B1-B4 (see Methods).

\subsection{High-fidelity single-qubit gates}\label{result:high_fid_1q}
\begin{figure}[H]
\centering
\captionsetup{font=footnotesize,skip=0pt,width=1\linewidth}
\includegraphics[width=1\textwidth]{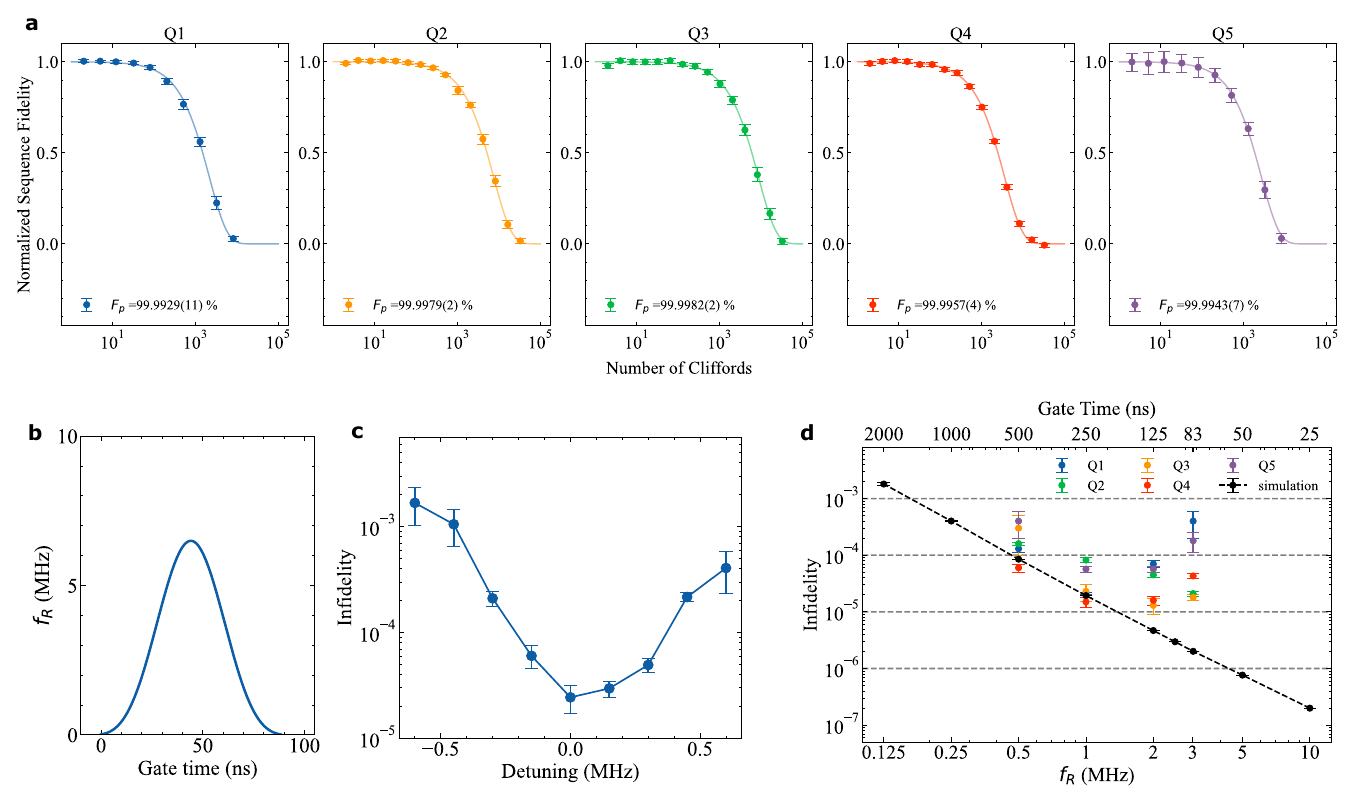}
\caption{
\textbf{High-fidelity single-qubit gates.}
\textbf{a} Single-qubit RB experiments with individually addressed qubits using Kaiser-window pulses. All five qubits reach primitive $\pi/2$ gate fidelities $F_p$ above $99.99$\%. The gate times of the primitive gates for the five qubits are 125~ns, 83~ns, 83~ns, 83~ns, and 125~ns, respectively. Error bars on the data points represent the 95\% confidence interval calculated from the standard deviation of the data. Error in fidelity is calculated from the fitting covariance matrix.
\textbf{b} Kaiser-window control pulse used in the RB experiments.
\textbf{c} RB fidelities under MW frequency detuning. RB measurements show fidelities $>99.99$\% over a range of $\sim 0.4$~MHz. The slight skew toward positive detuning is attributed to MW-induced heating (see Extended Fig.~\ref{extend:heating_shift} and Supplementary Materials). Errors are derived from the fitting covariance matrix.
 \textbf{d} Single-qubit randomized benchmarking (RB) at varying operation speeds using Kaiser-window pulses. The operation speed is defined as the Rabi frequency \( f_{\text{R}} \) of a square pulse with the same area as the applied pulse. Simulations use noise extracted from independent measurements of qubits under idling conditions (Extended Data Fig.~\ref{extend:simulations}), which defines a lower bound set by the dephasing limit. At faster gate speeds, the deviation of the experimental data from the simulations results from MW-induced decoherence caused by the control pulses, not captured in the simulations. These effects become more pronounced at higher operation speeds, since faster gates require stronger MW drive.
 }
\label{fig2}
\end{figure}

We first operate each qubit individually. Single-qubit RB experiments are performed on all five qubits using Kaiser-window control pulses, shown in Fig.~\ref{fig2} (b). The Kaiser-window pulse features an excitation spectrum engineered to suppress direct MW-drive-induced qubit–qubit crosstalk from spectral leakage ~\cite{rimbach-russ_simple_2023}, which is essential for spin qubit arrays operated with a shared control line. A key requirement for achieving high-fidelity control with tailored pulses such as the Kaiser window is to operate the qubits in a regime where the Rabi frequency scales linearly with the drive amplitude. To verify this, we measure the Rabi frequency $f_{\text{R}}$ as a function of drive amplitude for all five qubits and confirm linearity up to $f_{\text{R}} \sim 8$~MHz, as shown in Fig.\ref{fig1}(h). Beyond 9~MHz, qubit Q$_{1}$ exhibits reduced driven coherence time $T_{2}^{\text{R}}$. Based on this observation and additional considerations from the tailored pulses, we choose a minimal gate time of $t_{\text{g}} = 83$~ns.

The Clifford gate set is composed solely of $X$ and $Y$ gates, each corresponding to a $\pi/2$ pulse, such that the fitted primitive gate fidelity matches the fidelity of the single-qubit $\pi/2$ gate (Methods). To minimize qubit idling, where they are more susceptible to dephasing noise, we replace the identity gate $I$ in the gate set with $4 \times X$ or $0 \times X$. Both replacements result in similar fidelities, enhancing the robustness of the gate sequences against dephasing \cite{xue_benchmarking_2019}. For the same reason, the idle time between primitive quantum gates is minimized to 2~ns, the hardware limit of our control equipment (see Methods). The qubit frequency and pulse amplitude for the single-qubit gates are calibrated and fine-tuned before the experiment (Extended Data Fig.~\ref{extend:gate_calib}). Following these calibrations, we measure the RB decay for all five qubits, as shown in Fig.~\ref{fig2}(a). All qubits reach primitive single-qubit fidelities above $99.99$\%, with some approaching $99.999$\%. Interleaved RB \cite{magesan_efficient_2012} confirms that the $X$ and $Y$ gates perform similarly, both achieving fidelities well above 99.99\% (Extended Data Fig.~\ref{extend:inter_rb}). 

Next, we test the robustness of our high-fidelity control against frequency detuning by measuring the RB fidelity of qubit Q$_3$ as a function of MW-frequency offset, as shown in Fig.~\ref{fig2}(c). We find that fidelities remain above $99.99$\% across a detuning window of $\sim 0.4$ MHz. Notably, this window is larger than the typical qubit frequency fluctuations in our devices ($\sim 0.1$~MHz), which are primarily caused by charge noise \cite{yoneda_quantum-dot_2018, RojasArias2023, noiri_fast_2022}.

Gate fidelity depends on a trade-off between dephasing at low speeds and reduced driven coherence at high speeds. We evaluate this behavior by performing single-qubit RB experiments at varying operation speeds (Fig.~\ref{fig2}{(d)}).  The resulting infidelity increases both for slow gates, due to charge-noise-induced dephasing~\cite{yoneda_noise-correlation_2023}, and for fast gates due to additional MW-induced decoherence ~\cite{takeda_fault-tolerant_2016, yoneda_quantum-dot_2018, vallabhapurapu_fast_2021}. This trade-off is reflected in the qubit operation quality factors, which exhibit a maximum at the optimal driving speed (Extended Data Fig.~\ref{extend:qubit_power_dep}). For qubits Q$_{1}$ to Q$_{4}$, the optimal gate time results in a fidelity close to 99.999\%, while Q$_{5}$ reaches an optimal fidelity of $\sim 99.995\%$, consistent with its lower maximum quality factor. We attribute the variations between qubits to differences in the confinement potential, likely arising from local fabrication, tuning conditions, and material variability common in academic devices such as the one used here~\cite{Pioro-Ladriere_2008, malkoc_optimal_2016}.

To quantitatively assess these trends, we compare the experimental fidelities to numerical simulations based on dephasing noise measured under idling conditions (Extended Data Fig.~\ref{extend:simulations}). As expected, the simulations accurately reproduce the infidelity trends at slower gate speeds (500~ns-250~ns), consistent with the expected infidelity scaling \(1 - F \propto t_{\text{g}}^2\) for dephasing under quasi-static noise (see Methods for details). However, at faster gate speeds, the experimental fidelities are lower than predicted. This indicates that MW-induced noise from qubit operations, absent during idling and thus not included in the model, emerges at faster speeds and contributes to decoherence. The associated infidelity can be reproduced by modeling a moderate increase in the qubit noise level caused by the applied MW signals (Extended Data Fig.~\ref{extend:simulations}).

\subsection{Simultaneous operation of two qubits}\label{result:simultaneous_drive}
 As a first demonstration of high-fidelity parallel control, we investigate simultaneous operation of two qubits. We benchmark the fidelity of this parallel control using a simultaneous randomized benchmarking (SRB) experiment on qubits Q$_3$ and Q$_4$ \cite{gambetta_characterization_2012}. In this experiment, independent randomized Clifford sequences are applied simultaneously to both qubits, and the resulting decays are fitted to extract individual gate fidelities under simultaneous operation (Methods). Simultaneous quantum gates are implemented by combining the single-qubit MW pulses for each qubit using a MW combiner, enabling operation of multiple qubits with a shared control line (see also Methods).
Because of the AC-Stark effect, MW control pulses addressed to one qubit also expose neighboring qubits on the shared control line to off-resonant fields, shifting their resonance frequencies and inducing $z$-phase errors (Extended Data Fig.~\ref{extend:crosstalk_phase}).
These crosstalk phase shifts are quadratic in the off-resonant drive amplitude, with the sign opposite for each qubit in the pair, as is characteristic of the AC-Stark effect. Simultaneous operations also modify the optimal MW amplitudes required for $\pi/2$ rotations, introducing $x$-rotation errors due to slight variations in electron displacement from the combined MW drives. To achieve high-fidelity simultaneous operations, both the $x$-rotation and $z$-phase errors must be calibrated and compensated.

\begin{figure}[H]
\centering
\captionsetup{font=footnotesize,skip=0pt,width=1\linewidth}
\includegraphics[width=1\textwidth]{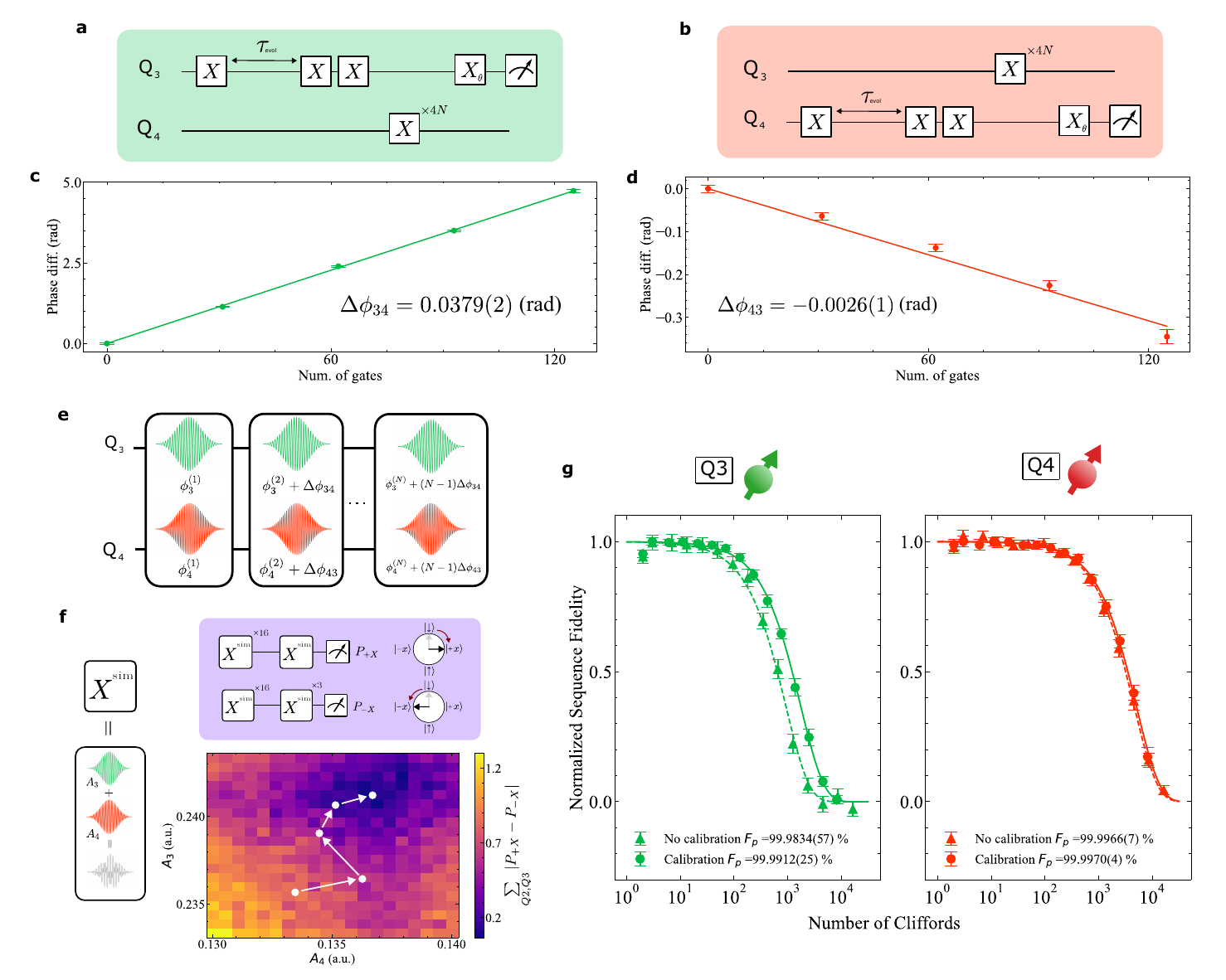}
\caption{\textbf{Crosstalk phase in simultaneous two-qubit operation.} 
\textbf{a, b} Sequence for measuring the crosstalk phase errors.
\textbf{c} Measured crosstalk phase errors accumulated on qubit Q$_{3}$ when operating qubit Q$_{4}$ and \textbf{d} on qubit Q$_{4}$ when operating qubit Q$_{3}$.
\textbf{e} Compensation of crosstalk phase error during simultaneous operations. For each pulse, the MW phase in the following gates on qubit Q$_{3}$ (Q$_{4}$) is shifted by $\Delta\phi_{34}$ ($\Delta\phi_{43}$).
\textbf{f} Calibration of the operation amplitudes A$_{3}$ and A$_{4}$ for simultaneous operations on Q$_{3}$ and Q$_{4}$. The simultaneous $X$ gate is implemented by combining the individual qubit driving signals and applying them to the MW gate.
 {\textbf{g} SRB decays of qubit Q$_{3}$ and Q$_{4}$ without and with the phase and amplitude calibration. With calibration, Q$_{3}$ shows a significant fidelity increase, while Q$_{4}$ remains mostly unaffected, due to a larger crosstalk phase error in Q$_{3}$.}
}\label{fig3}
\end{figure}

To calibrate the crosstalk phase errors, we apply a Hahn-echo sequence to the target qubit while performing $4N$ $X$ gates on another qubit (Fig.\ref{fig3} (a, b)) and measure the resulting phase accumulation per gate (Fig.\ref{fig3} (c, d)). The crosstalk phase accumulated on qubit Q$_{i}$ when operating qubit Q$_{j}$ is denoted as $\Delta\phi_{ij}$. For $Q_{3}$ and $Q_{4}$, the measured crosstalk phases are $\Delta \phi_{34}=0.0379(2)$ rad and $\Delta \phi_{43}=-0.0026(1)$ rad.  {The discrepancy in phase shift magnitudes arises from the different power required to drive each qubit at the same operation speed (Extended Data Fig.~\ref{extend:qubit_power_dep}): Q$_{3}$ requires less power than Q$_{4}$, resulting in a smaller off-resonant drive and, consequently, a smaller AC-Stark shift experienced by Q$_{4}$.} These phases are compensated during the simultaneous pulse sequence by offsetting the MW phases during the experiment, as illustrated in Fig.~\ref{fig3} (e).

After measuring the crosstalk phase errors, we calibrate the operation pulse amplitudes, adjusting $A_{3}$ and $A_{4}$ for the two-qubit simultaneous operation (Fig.\ref{fig3} (f)). We apply 16 simultaneous $X$ gates to the qubit pairs, followed by one (three) simultaneous $X$ gate(s), which flip the qubits into $|+x,+x\rangle$ ($|-x,-x\rangle$) state to maximize sensitivity to over- or under-rotation errors. Since the probabilities $P_{+X}$ and $P_{-X}$ for the two sequences are equal at an optimal MW amplitude, we use an optimizer to minimize the summed probability difference $\sum_{Q_{3},Q_{4}}|P_{+X}-P_{-X}|$ of both qubits to obtain the optimal pulse amplitudes. Fig.~\ref{fig3} (f) shows a 2D map of the probability difference, along with an illustration of the optimization process. 

 {To demonstrate the effect of our calibration scheme, we perform a two-qubit SRB experiment on qubits Q$_3$ and Q$_4$, both with and without applying the calibrations for the simultaneous operations (Fig.~\ref{fig3} (g)). In the uncalibrated case, the individual single-qubit gate pulses are summed and applied to qubit Q$_{3}$ and Q$_{4}$ directly. Without calibration, Q$_{4}$ still achieves a gate fidelity above 99.99\%, while Q$_{3}$ remains limited to 99.9\% due to a sizable uncorrected phase accumulation $\Delta \phi_{34}$. After applying the calibrations, both qubits reach fidelities exceeding 99.99\%.}

\subsection{Simultaneous operation of multiple qubits}\label{result:multi_qubit_drive}
\begin{figure}[H]
\centering
\captionsetup{font=footnotesize,skip=0pt,width=1\linewidth}
\includegraphics[width=1\textwidth]{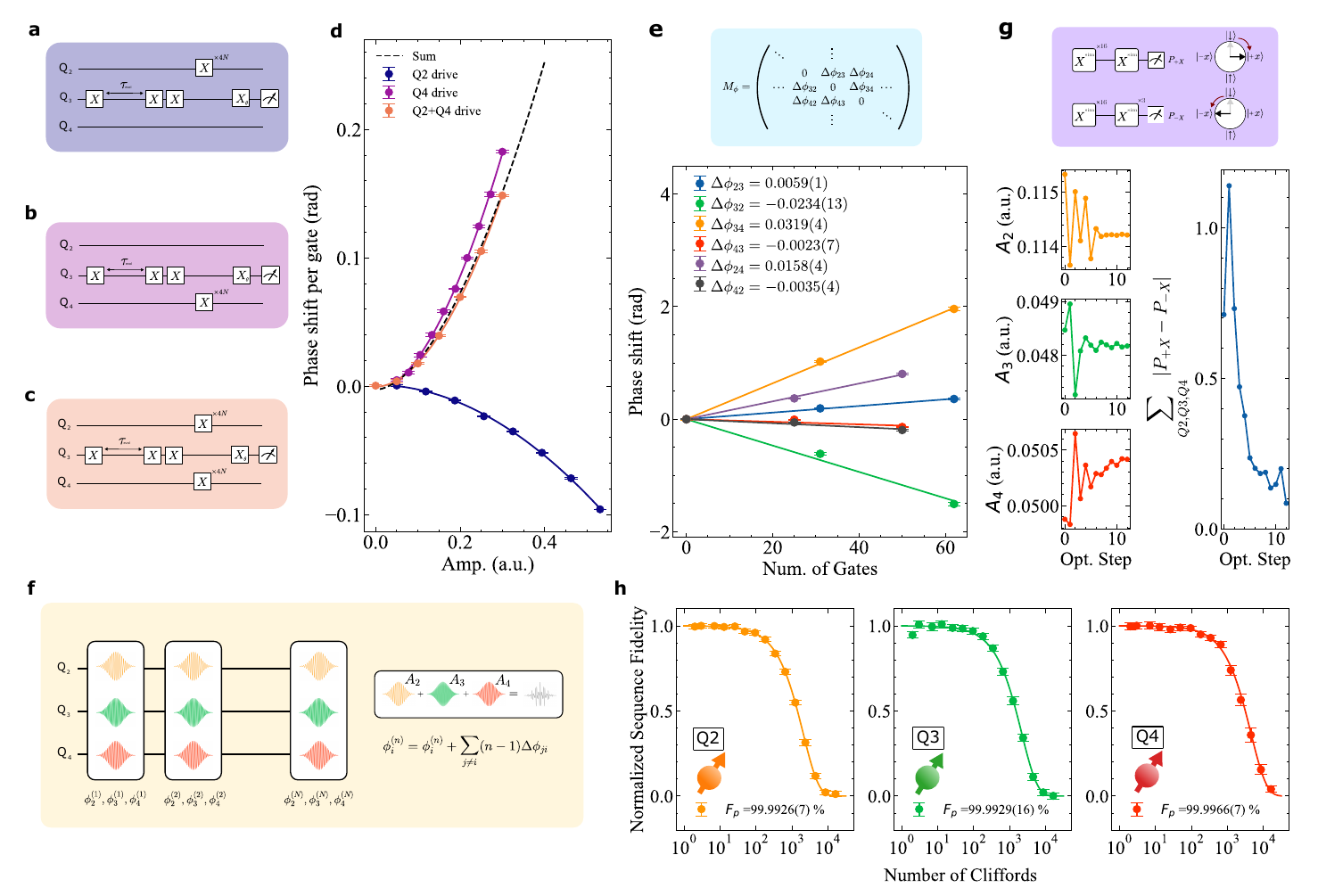}
\caption{\textbf{Generalized calibration of parallel multi-qubit operations.} 
\textbf{a-c} Sequence for measuring crosstalk phase error accumulated on qubit Q$_{3}$ while operating qubit Q$_{2}$, qubit Q$_{4}$ and both Q$_{2}$ and Q$_{4}$ simultaneously.
\textbf{d} Amplitude dependence of the crosstalk phases. We fit the phase accumulations with $A V_{\text{amp}}^{\alpha}$ and repeat the experiment for other qubit pairs. We find an average $\alpha = 1.91(5)$, in good agreement with the AC-Stark shift (Extended Data Fig.~\ref{extend:crosstalk_phase}). The black dashed line shows the sum of the two fits in the individual operation cases.
\textbf{e} Crosstalk phase measured for all qubit pairs in a qubit Q$_{2,3,4}$ three-qubit simultaneous drive. 
\textbf{f} {Illustration of the protocol to compensate for the crosstalk phase error for the multi-qubit simultaneous operations.} 
\textbf{g} Amplitude calibration for a three-qubit simultaneous operation.
\textbf{h} Three-qubit SRB decays showing >99.99\% primitive gate fidelity. 
}\label{fig4}
\end{figure}

We now extend our parallel-control experiments to multiple qubits in the array. The crosstalk phases are calibrated for multi-qubit operations by measuring the phases accumulated on each qubit individually and summing these contributions linearly. 
To validate this approach, we measure the crosstalk phases accumulated on Q$_{3}$ when operating Q$_{2}$ or Q$_{4}$ individually (see Fig.~\ref{fig4}(a, b) for the quantum circuit) and compare them to simultaneous pulses on both qubits (Fig.~\ref{fig4}(c)). The measured phases as a function of control pulse amplitude are shown in Fig.~\ref{fig4}(d).
Summing the fitted individual-phase results (black dashed line) agrees well with simultaneous pulse measurements, demonstrating that crosstalk phases can be measured pairwise and summed linearly. This method enables precise, scalable compensation of phase crosstalk in multi-qubit control sequences.

Fig.~\ref{fig4} (e, f) illustrates the calibration procedure for general multi-qubit simultaneous operations, demonstrated here for the three-qubit case with qubits Q$_{2}$, Q$_{3}$, and Q$_{4}$. The simultaneous $\pi/2$ gate duration is set to 250~ns to reduce the peak driving amplitude. Crosstalk phases $\Delta\phi_{ij}$ between each qubit pair are measured individually and compensated within the gate sequences by shifting the MW phase on each qubit Q$_{i}$ by $\sum_{j\neq i}\Delta\phi_{ij}$, as illustrated in Fig.~\ref{fig4}{(f)}. The simultaneous operation amplitudes are then calibrated using the same method described for the two-qubit case. Fig.\ref{fig4}{(g)} illustrates the three-qubit optimization process, showing the driving amplitudes $A_{2}$, $A_{3}$, and $A_{4}$, and the resulting probability difference at each step. After approximately 10~steps, the amplitudes converge, minimizing the probability difference. Using these calibrated phases and amplitudes, we perform a three-qubit SRB experiment and achieve primitive $\pi/2$ gate fidelities >99.99\% for all three qubits (Fig.~\ref{fig4}(h)).

\begin{figure}[H]
\centering
\captionsetup{font=footnotesize,skip=0pt,width=1\linewidth}
\includegraphics[width=1\textwidth]{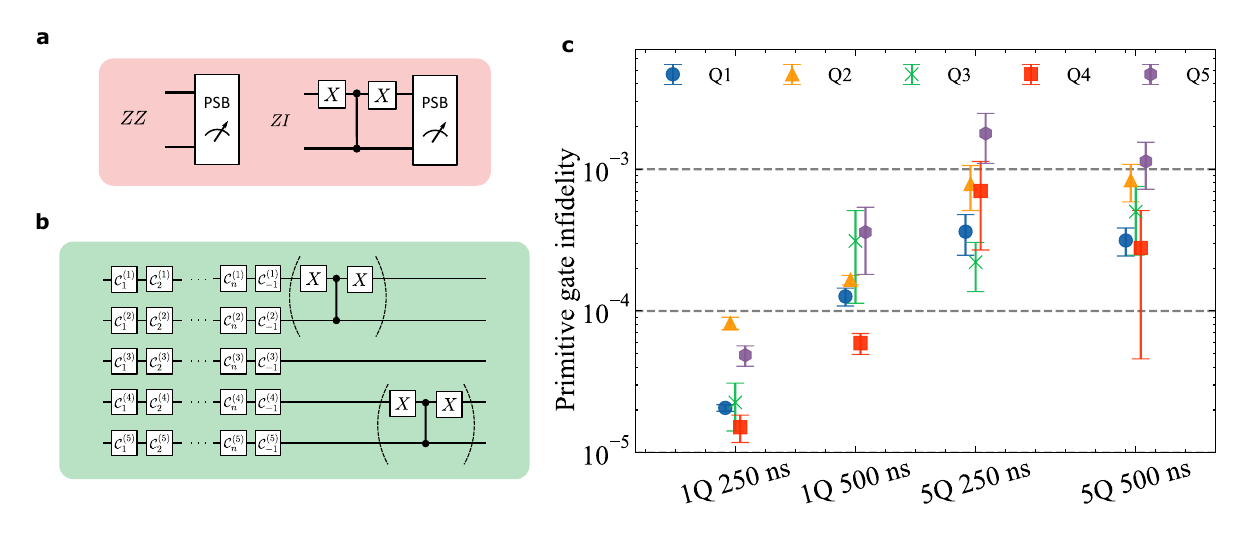}
\caption{\textbf{Five-qubit simultaneous operation.} 
\textbf{a} Tomographic readout protocol used to obtain the individual spin states from the PSB pairs Q$_{1,2}$ and Q$_{4,5}$. 
\textbf{b} Five-qubit SRB sequence with tomographic readout. The sequence is repeated with ZZ and ZI measurements on PSB pairs Q$_{1,2}$ and Q$_{4,5}$ to reconstruct all individual qubit states.
\textbf{c}  {Primitive gate infidelity of single-qubit operation and five-qubit simultaneous operation with $X$ gate times of 250~ns and 500~ns. Four of five qubits exceed fidelities of 99.9\%, with Q$_5$ reaching approximately 99.9\%.}
}\label{fig5}
\end{figure}

Finally, we explore the limits of parallel control with our shared control line approach by operating all five qubits simultaneously and discuss the implications for larger arrays. To achieve this, we calibrate all crosstalk phases and control pulse amplitudes using the pairwise-calibration procedure described above. We also compare this pairwise calibration approach with a full calibration, in which the phase accumulation on each qubit is measured while the other four are operated, and observe no difference in performance. This underscores that the pairwise approach enables scalable and resource-efficient parallel control across multiple qubits. Due to readout limitations (only three bits of information per cycle), we perform tomographic readout \cite{philips_universal_2022} on the qubit pairs Q$_{1,2}$ and Q$_{4,5}$ to reconstruct the individual qubit states (see Fig.~\ref{fig5}(a) and Methods). The complete five-qubit SRB sequence is illustrated in Fig.~\ref{fig5}(b), where each random Clifford sequence is repeated twice to perform the two required tomographic measurements ($ZZ$ and $ZI$) needed to reconstruct the full five-qubit state.

We perform SRB experiments with gate durations of $t_{\text{g}}=$250~ns and 500~ns (Fig.\ref{fig5}(c)). In both cases, overall fidelities remain similar but lower than in the three-qubit parallel operation, with four out of five qubits achieving fidelities above 99.9\%. This behavior reflects opposing error mechanisms: at $ t_{\text{g}} =$250~ns, the higher total MW power required for simultaneous operation increases decoherence, likely due to heating or MW-induced noise, thereby reducing performance relative to single-qubit gates. At $t_{\text{g}}=$500~ns, the lower drive amplitudes mitigate these effects. Still, the longer gate duration introduces increased dephasing from low-frequency noise, yielding fidelities comparable to single-qubit operation (see also Extended Data Table\ref{extend:all_fid}). The similar fidelities observed at $t_{\text{g}}=$250~ns and 500~ns indicate that high-fidelity simultaneous control with $\sim$99.9\% fidelities could be extended to ten qubits at $t_{\text{g}}=$500~ns, where the total MW power matches that of five-qubit operation at $t_{\text{g}}=$250~ns.

\section{Conclusions}\label{dicussion}
We demonstrate high-fidelity (>99.99\%) single-qubit control across all five qubits in a silicon spin qubit array using tailored Kaiser-window control pulses and maintain these fidelities during simultaneous operation of up to three qubits through a shared control line. Extending to simultaneous control of all five qubits, we achieve fidelities at the practical fault-tolerant threshold of 99.9\%, and argue that this control scheme can be extended to ten qubits in the current device architecture. Together, these results demonstrate that high-fidelity parallel control across multiple silicon spin qubits is achievable, addressing a central challenge in scaling spin-based quantum processors.

This level of parallel control is enabled by a scalable calibration protocol that compensates for MW-induced, qubit-specific phase shifts from off-resonant AC Stark shifts. For simultaneous control of $N$ qubits, only $N(N-1)$ pairwise calibrations are required, avoiding the exponential overhead of full crosstalk characterization. In general, the magnitude of these phase shifts scales inversely with detuning, such that corrections become negligible for sufficiently detuned qubit pairs (Extended Data Fig.~\ref{extend:crosstalk_phase}), further reducing the calibration overhead in larger arrays. 

While fidelities remain high, simultaneous five-qubit operation is currently limited by MW-induced decoherence at fast gate speeds and by dephasing at longer gate durations. Modest improvements to the micromagnet design can double the driving-field gradient (see Extended Data Fig.~\ref{extend:micromagnet_sim} and Supplementary Materials), enabling faster gates at reduced power. This would mitigate MW-induced decoherence and extend the dephasing-limited regime to higher speeds, allowing fidelities above 99.999\% at modest gate speeds of 2–3~MHz (see Fig.~2d) and supporting high-fidelity parallel control of larger qubit numbers.

To further enhance control performance and scalability, the micromagnet can be replaced by nanomagnets integrated into the gate layer or back-end-of-line wiring. Placing magnetic structures closer to the qubits provides stronger local field gradients \cite{Tadokoro2021_twoDsiarray, Aldeghi2025_nanomagnets} and preserves gradient strength in multi-layer routing architectures required for two- or quasi-two-dimensional spin qubit arrays \cite{Li_trilinear, veldhorst_silicon_2017}. Without such integration, the increased vertical separation between qubits and magnetic structures above the routing layers will otherwise degrade control performance \cite{George2025-12qarrays, weinstein_universal_2023}, limiting the scalability of parallel qubit control.

Our shared control approach directly addresses a key scaling bottleneck in silicon spin qubits, enabling parallel and high-fidelity operation. Driving multiple qubits from a single impedance-matched control line reduces wiring overhead and simplifies on-chip control infrastructure. Combined with compact pairwise calibration and targeted device improvements, this offers a practical control strategy for scalable silicon qubit unit cells. Crucially, the shared-control-line architecture can be extended beyond local units to support the parallel control of multiple subarrays, with total MW drive power and calibratable AC Stark shifts as the primary constraints. Together, these advances define a realistic path toward scalable, parallel qubit control in silicon, which is a core requirement for the development of large-scale quantum processors.

\section*{Data Availability}
All data from this study will be made available in a Zenodo repository.

\section*{Acknowledgments}
We thank M.~Rimbach-Russ and L.~M.~K. Vandersypen for helpful discussions, and J.~S.~Rojas-Arias for assistance with the RB model. We also acknowledge R.~Kuroda for support with sample fabrication. This work was supported financially by Core Research for Evolutional Science and Technology (CREST), Japan Science and Technology Agency (JST)  (JPMJCR1675), MEXT Quantum Leap Flagship Program (MEXT Q-LEAP) grant numbers JPMXS0118069228, JST Moonshot R\&D Grant Number JPMJMS226B, and JSPS KAKENHI grant numbers 18H01819 and 20H00237. T.N. acknowledges support from JST PRESTO grant number JPMJPR2017. A.N. acknowledges support from JST PRESTO Grant Number JPMJPR23F8. Y.H.W. acknowledges support by RIKEN's IPA program. H.-S. Goan acknowledges support from the National Science and Technology Council, Taiwan, under Grants No. NSTC 113-2112-M-002-022-MY3, No. NSTC 113-2119-M-002-021, No. 114-2119-M-002-018, No. NSTC 114-2119-M-002-017-MY3, from the US Air Force Office of Scientific Research under Award Number FA2386-23-1-4052 and from the National Taiwan University under  Grants No. NTU-CC-114L8950, No. NTU-CC114L895004 and No. NTU-CC-114L8517. H.-S. Goan. is also grateful for the support of the "Center for Advanced Computing and Imaging in Biomedicine (NTU-114L900702)" through the Featured Areas Research Center Program within the framework of the Higher Education Sprout Project by the Ministry of Education (MOE), Taiwan, the support of Taiwan Semiconductor Research Institute (TSRI) through the Joint Developed Project (JDP) and the support from the Physics Division, National Center for Theoretical Sciences, Taiwan.

\section*{Author Contribution}
Y.H.W., L.C.C., and P.B. performed the experiment. A.N. fabricated the device. I.K.J., K.T., A.N.,  T.K., T.N., and H.-S.G. contributed to the data acquisition and discussed the results. G.S. developed and supplied the silicon-28/silicon-germanium heterostructure. Y.-H.W. and L.C.C. wrote the manuscript with inputs from all co-authors. S.T. supervised the project.

\section*{Declarations}
The authors declare no competing interests.

\section*{Additional Information}
\textbf{Correspondence and requests for materials} should be addressed to L.C.C. (email: leon.camenzind@riken.jp), Y.-H.W. (email:yi-hsien.wu@a.riken.jp) or S.T. (email: tarucha@riken.jp).

\bibliography{template_bib}
\bibliographystyle{sn-standardnature_nourl}

\section*{Methods}\label{methods}

\subsection*{Device fabrication}\label{methods:device}
The quantum dots were defined in a 9~nm thick isotopically enriched silicon quantum well with a residual Si-29 concentration of 800~ppm. The quantum well is covered by a 30 nm SiGe buffer layer and approximately 2 nm of silicon oxide. A 15~nm Al$_{2}$O$_{3}$ layer was then deposited on the wafer using atomic layer deposition (ALD). Three layers of overlapping aluminum gates were fabricated through electron-beam lithography and lift-off processes. The aluminum was oxidized, and the resulting native aluminum oxide serves as an insulator between the gate layers. A micromagnet was fabricated on top of the device to provide magnetic field gradients for qubit addressability and operations. The micromagnet consists of a stack of 5~nm titanium and 250~nm cobalt films and is positioned on top of the overlapping gates, separated by a 30~nm ALD aluminum oxide layer. The micromagnet design is similar to those used in our previous devices\cite{noiri_fast_2022, takeda_fault-tolerant_2016, yoneda_robust_2015} with a gap size of 400~nm, see also Extended Fig.~\ref{extend:micromagnet_sim} and Supplementary Materials for more details.

\subsection*{Qubit control}\label{methods:qubit_ctrl}
To control the qubit energies, quantum dot confinements, and tunnel couplings, we apply DC voltages using D5a digital-to-analog converters (DACs) from Qblox and employ baseband pulses generated by Keysight M3201A arbitrary waveform generators (AWGs) to pulse the qubits between operation and readout points. Qubit control pulses are generated by Keysight M3202A arbitrary waveform generators (AWGs) and mixed with local oscillators (LO) from a multi-channel signal generator (APMS20G4) using IQ mixers. The single-qubit operations of the five qubits are facilitated through three IQ mixers (2xMMIQ-0626HS and 1xMLIQ02181). The IQ mixers have been carefully calibrated to minimize leakage of local oscillator and image signals. The resulting signals are combined in a power splitter (ZC8PD-5R263-S+), amplified (ZVA-183G-S+), and bandpass-filtered. An MW switch (P9402C) suppresses MW leakage outside the operation stage. The MW signal is then routed through a low-loss superconducting NbTi coaxial line inside the dilution refrigerator, entering a designated impedance-controlled MW line on the sample PCB. From there, the signal is transmitted via bond wires to the quantum chip, reaching the designated MW gate (see Fig. 1a).

 {To perform the CROT and CZ two-qubit gates required for initializing, operating, and reading out the quantum processor, the exchange interaction between neighboring qubits is increased from the residual exchange 
 $J < 16.6$ kHz (Extended Data Fig.~\ref{extend:res_exchange}) to a few MHz using baseband pulses. In this device, spin-valley mixing occurs as the exchange interaction is increased, limiting the fidelity of our CZ two-qubit gates to $<$97\%. Under single-qubit operation conditions, where only residual exchange is present, the device can be tuned to eliminate the spin-valley mixing effect across all five qubits.}

\subsection*{Kaiser-window pulse} \label{methods:pulse_shaping}
We use the Kaiser-window envelope, defined as
\begin{equation}
w(t) = 
\begin{cases}
\frac{I_0 \left( \beta \sqrt{1 - \left( \frac{t - t_{\text{g}}/2}{t_{\text{g}}/2} \right)^2} \right)}{I_0(\beta)}, & 0 \leq t \leq t_{\text{g}} \\
0, & \text{otherwise}
\end{cases}
\end{equation}
where $I_0$ is the zeroth-order modified Bessel function of the first kind and $\beta$ is the pulse parameter. In our experiments, we optimized and chose $\beta=8$ such that the maximal driving speed does not cause the loss of qubit coherence (Extended Data Fig.~\ref{extend:qubit_power_dep}). We convert from a rectangular (square) pulse of amplitude $V_{\text{rec}}$ and gate time $t_{\text{g}}$ to a Kaiser pulse of $V_{\text{Kaiser}}w(t)$ of the same gate time. The Kaiser pulse amplitude is calculated with $V_{\text{Kaiser}} = V_{\text{rec}} t_{\text{g}}/\mathcal{I}$ where $\mathcal{I}$ is the integral of the Kaiser window $\mathcal{I} = \int_{0}^{t_{\text{g}}} w(t) dt$.

\subsection*{Randomized benchmarking} \label{methods:rb}
The Clifford gate set used in the RB experiments includes only $X$ and $Y$ gates, allowing the primitive gate fidelity to be directly attributed to the fidelity of the corresponding  $X$ and $Y$ gates, each representing a $\pi/2$ rotation. The full Clifford set decomposition is $\{ XXXX, Y, X, YX, XY, YY, XX, YXX, YYX, YXY, YYY, \\ XXX, XXY, XYY, YXYY, YYXY, XXXY, YXXX, YYXX, XYYY, YYYX, YYYXY, \\ YXXXY, YXYYY \}$ where the identity gate is replaced with four $X$ gates to prevent qubit idling. The Clifford gate set comprises 78 primitive gates when including the $4\times X$ identity replacement and 74 when using the $0\times X$ replacement. This corresponds to an average of 3.25 and 3.217 primitive gates per Clifford, respectively. In the applied sequences, the primitive gates are interleaved with 2~ns of idle time, corresponding to the minimum delay allowed by the FPGA for updating the MW phase.

We chose randomized sequences with a maximal Clifford gate length of either $2^{10}=1,024, 2^{13}=8,192, 2^{14}=16,384$, or $2^{15}=32,768$ for the single-qubit and the simultaneous RB experiments. For each gate length, we randomize 35 times and average over 350 single-shot measurements in each randomization to obtain the spin state probabilities. At the end of each sequence, we apply a recovery gate that ensures the entire sequence acts as either an identity operation or a $X^{2}$ flip. Each sequence then gives the corresponding flip probability $F_{\text{noflip}}(n)$ and $F_{\text{flip}}(n)$. We fit the difference $F(n)=F_{\text{flip}}(n)-F_{\text{noflip}}(n)$ with an exponential function $F(n)=Ap^{n}$ to obtain the depolarizing parameter $p$. We calculate the Clifford fidelity with $F_{C} = (1+p)/2$ and obtain the primitive gate fidelity by $F_{p} = 1 - (1-F_{C})/3.25$ \cite{lawrie_simultaneous_2023}.

Simultaneous randomized benchmarking is performed by implementing single-qubit RB experiments in parallel on multiple qubits. For each qubit, distinct randomized benchmarking sequences are generated, each composed of $\pi/2$ gates with identical gate times. These sequences are temporally synchronized, enabling a synchronized operation cycle with simultaneous operations across multiple qubits. Due to the generation of different randomized benchmarking (RB) sequences, the number of primitive gates in each sequence may vary across qubits. Any additional primitive gates are applied either through simultaneous pulses on a subset of qubits or via individual single-qubit pulses. The difference in the number of primitive gates among individual qubit sequences can be as large as 200 for a sequence containing 10,000 Clifford gates. 

Each qubit's sequence is analyzed independently, similar to a single-qubit randomized benchmarking (RB) experiment, to determine the fidelity of individual single-qubit primitive gates. This separated analysis helps identify qubit-specific errors. A joint fidelity is obtained by multiplying the individual fidelities.

\subsection*{Five-qubit tomographic readout} \label{methods:tomo_readout}
The five-qubit operation cycle provides only 3 out of 5 bits of information (2 PSB and 1 QND readout). Therefore, a tomographic readout is required to fully measure the five-qubit spin state in the five-qubit SRB experiment. To achieve this, we repeat each operation sequence twice: one $ZZ$ component read out and one $ZI$ component read out. The ZZ component is measured directly with the ZZ measurement operator of the PSB readout. The ZI component is obtained by mapping the ZZ operator to a ZI operator by applying a CNOT gate before the PSB readout. We synthesize the CNOT using a CZ and single-qubit gates. The two-qubit state probabilities of qubit pairs Q$_{1,2}$ and Q$_{4,5}$ are used to construct single-qubit probabilities $P_{1,\uparrow}, P_{2,\uparrow}, P_{4,\uparrow}$ and $P_{5,\uparrow}$. We synthesize the CNOT gate using a CZ gate and a single-qubit gate \cite{mills_two-qubit_2022}.

By thresholding the two single-shot measurements $s_{ZZ}$ and $s_{ZI}$ with the readout threshold value $v_{\text{thres}}$, we obtain the two-qubit state probabilities 
\begin{align}
    P_{\uparrow\uparrow} & = (s_{ZZ}>v_{\text{thres}}) \& (s_{ZI}>v_{\text{thres}}), \\
    P_{\downarrow\uparrow} & = (s_{ZZ}<v_{\text{thres}}) \& (s_{ZI}<v_{\text{thres}}),\\
    P_{\uparrow\downarrow} & = (s_{ZZ}<v_{\text{thres}}) \& (s_{ZI}>v_{\text{thres}}), \\
    P_{\downarrow\downarrow} & = (s_{ZZ}>v_{\text{thres}}) \& (s_{ZI}<v_{\text{thres}}).
\end{align}
We then calculate the single-qubit state probability by
\begin{align}
    P_{1,\uparrow} &= P_{\uparrow\downarrow} + P_{\uparrow\uparrow}, \\
    P_{2,\uparrow} &= P_{\downarrow\uparrow} + P_{\uparrow\uparrow}.
\end{align}

Similarly, we obtain the probabilities $P_{4,\uparrow}$ and $P_{5,\uparrow}$. The sequence fidelity, which is fitted in the five-qubit SRB experiments, is calculated by $F_{\text{seq},i} = |P_{i,\uparrow}^{\text{flip}} - P_{i,\uparrow}^{\text{noflip}}|$. 

\subsection*{RB simulations incorporating measured qubit noise} \label{methods:simulations}
To model the impact of noise on the randomized benchmarking fidelities, we first extract a noise power spectral density (PSD) model from Ramsey and CPMG measurements performed under idling conditions, during which the quantum processor was not actively operated (Extended Data Fig.~\ref{extend:simulations}a). Based on this PSD, we generate noise contributions across three characteristic timescales determined by the experimental conditions: low-frequency (LF) noise (Extended Data Fig.~\ref{extend:simulations}b), intermediate-frequency (IF) noise, and high-frequency (HF) noise.

Randomized benchmarking (RB) sequences are constructed using the same gate set as in the experiments. Each primitive gate is simulated as a unitary time evolution $\exp(-i \hat{H}_\mathrm{op}(t) t_g)$, where $\hat{H}_\mathrm{op}(t)$ includes both the ideal drive and a $\sigma_z$ detuning~\cite{Kawakami2016_gatefidelityandcoherence} induced by noise:
\[
\hat{H}_\mathrm{op}(t) = 2\pi \left( \frac{\beta(t)}{2} \sigma_z + \frac{f_\mathrm{R}}{2} \left( \cos(\phi) \, \sigma_x + \sin(\phi) \, \sigma_y \right) \right),
\]
with $\beta(t) = \beta_\mathrm{LF} + \beta_\mathrm{IF} + \beta_\mathrm{HF}(t)$. 

Here, $\beta(t)$ describes the detuning noise in Hz. The noise is incorporated into the simulation as follows: the low-frequency noise component $\beta_\mathrm{LF}$ is held constant over each RB sequence and updated between sequences, while the intermediate-frequency component $\beta_\mathrm{IF}$ is updated after each primitive gate. To realistically capture correlations across timescales, LF and IF noise are generated jointly rather than independently. High-frequency noise $\beta_\mathrm{HF}(t)$ is modeled as a fluctuating detuning during gate execution, updated every 10~ns or 50~ns time step. Because HF noise has no measurable impact on the simulated fidelities at the experimentally relevant levels, it is neglected in the simulations.

After simulating the complete qubit evolution for an RB sequence under noise, we calculate the return probabilities $F_{\text{noflip}}$ and $F_{\text{flip}}$ and extract RB fidelities using the same procedure as for the experimental data.

The observed dependence of gate fidelity on gate duration can be understood analytically by considering the effect of slow detuning noise in the quasistatic regime. In this limit, the frequency detuning $\beta$ remains effectively constant during a gate of duration $t_g$, leading to a relative phase accumulation $\Delta \phi = 2\pi \beta t_g$ between the computational basis states \cite{Paladino2014_1fnoise_solidstate}. For small $\Delta \phi$, and assuming a symmetric distribution of detunings centered at zero, the $\pi/2$-gate infidelity becomes $1 - F = 1 - |\langle \psi_{\mathrm{ideal}} | \psi \rangle|^2 \propto \langle (\Delta \phi)^2 \rangle \propto \langle \beta^2 \rangle t_g^2$, linking gate error directly to the noise variance $\langle \beta^2 \rangle$ and the square of the gate duration \cite{Green2013_arb_control}. This quadratic dependence on $t_g$ is observed both in simulation and in experimental data where dephasing dominates ($t_g$= 250 and 500~ns), (see Fig.~\ref{fig2}d).

\section*{Extended Data}\label{extended_data}
\renewcommand{\figurename}{Extended data figure}
\renewcommand{\tablename}{Extended data table}
\setcounter{figure}{0} 

\begin{figure}[H]
\centering
\captionsetup{font=footnotesize,skip=0pt,width=1\linewidth}
\includegraphics[width=0.8\textwidth]{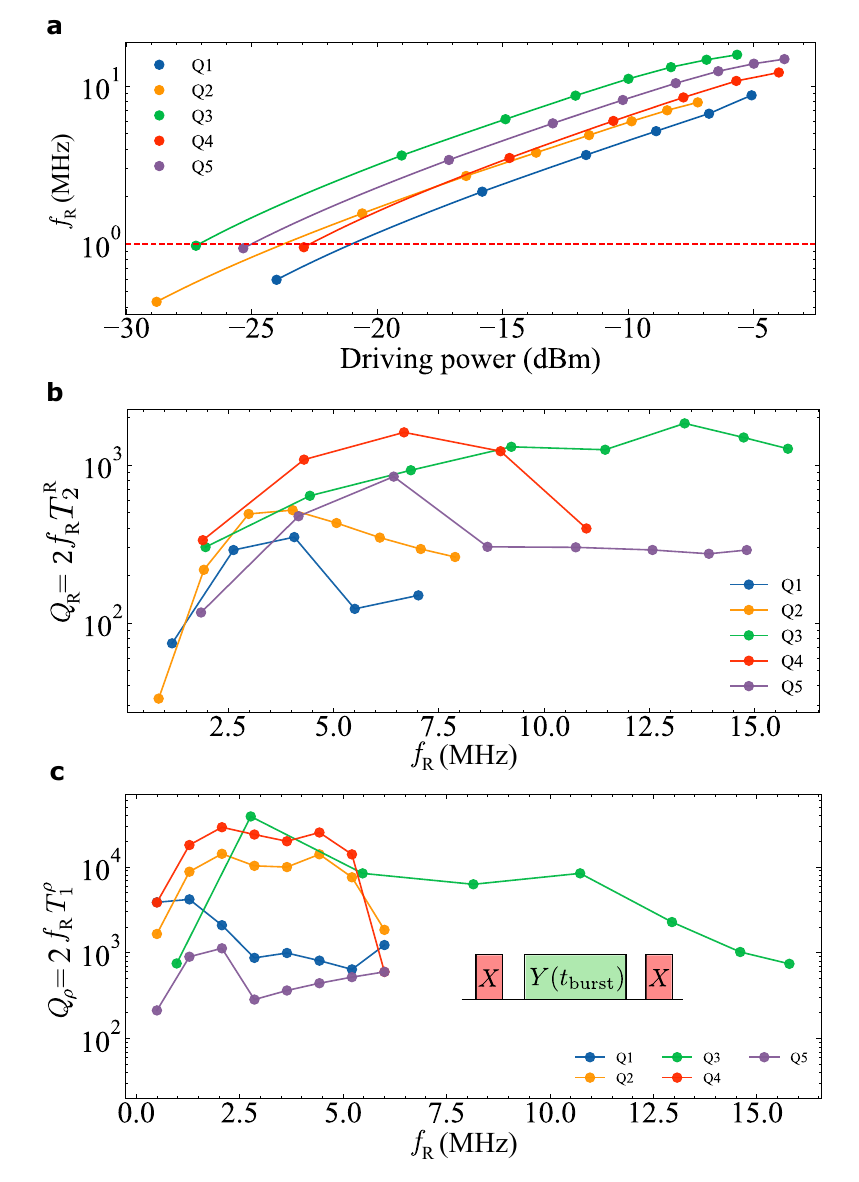}
\caption{\textbf{Qubit coherence when driven.} 
\textbf{a} Qubit Rabi frequency power dependence. The amplitude in main Fig.~\ref{fig1} (h) is converted into power measured by the spectrum analyzer.
\textbf{b} Rabi oscillation quality factors $Q_{\text{R}}=2f_{\text{R}}T_{2}^{\text{R}}$ under different driving Rabi frequency $f_{\text{R}}$. The Rabi coherence times are obtained by fitting the decays of Rabi oscillation amplitudes with $\exp{-t/T_{2}^{\text{R}}} W(t) $ where $W(t)=\left( 1+t^{2}/\left( f_{\text{R}} (T_{2}^{*})^{2} \right)^{2} \right)^{-1/4}$ takes into account the effect of dephasing \cite{nakajima_coherence_2020}.
\textbf{c} Spin-lock quality factor $Q_{\rho}=2f_{\text{R}}T_{1}^{\rho}$. The spin-lock decay time $T_{1}^{\rho}$ is measured by first flipping the spin to the $y-$axis with an X gate, followed by a drive in Y-direction of duration $t_{\text{burst}}$ and another X gate projecting the spin back to the principal axis.
}\label{extend:qubit_power_dep}
\end{figure}

\begin{figure}[H]
\centering
\captionsetup{font=footnotesize,skip=0pt,width=1\linewidth}
\includegraphics[width=1\textwidth]{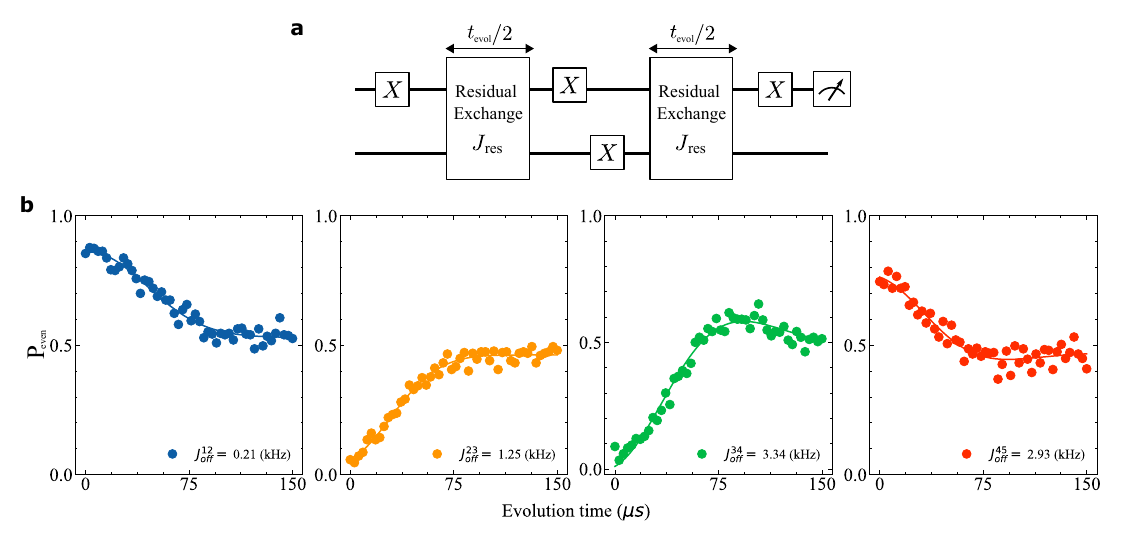}
\caption{\textbf{Residual exchange of the qubit pairs.} 
\textbf{a} Decoupled CZ (DCZ) sequence used to measure residual exchange $J_{\text{off}}$, defined as the unintended exchange interaction between qubits during single-qubit operations. This protocol isolates the effect of residual coupling by removing dynamic phase accumulations.
\textbf{b} Residual exchange measured in the four qubit pairs. In the presence of exchange, coherent oscillations would be expected. However, all measurements show no significant oscillations over the echo coherence time of $\sim 30$~$\mu$s. The observed decay is consistent with Hahn echo dephasing, setting an upper bound on the residual exchange of  $J_{\text{off}} < (1/30 \mu s)/2 \simeq 16.6$~kHz.
}\label{extend:res_exchange}
\end{figure}

\begin{figure}[H]
\centering
\captionsetup{font=footnotesize,skip=0pt,width=1\linewidth}
\includegraphics[width=1\textwidth]{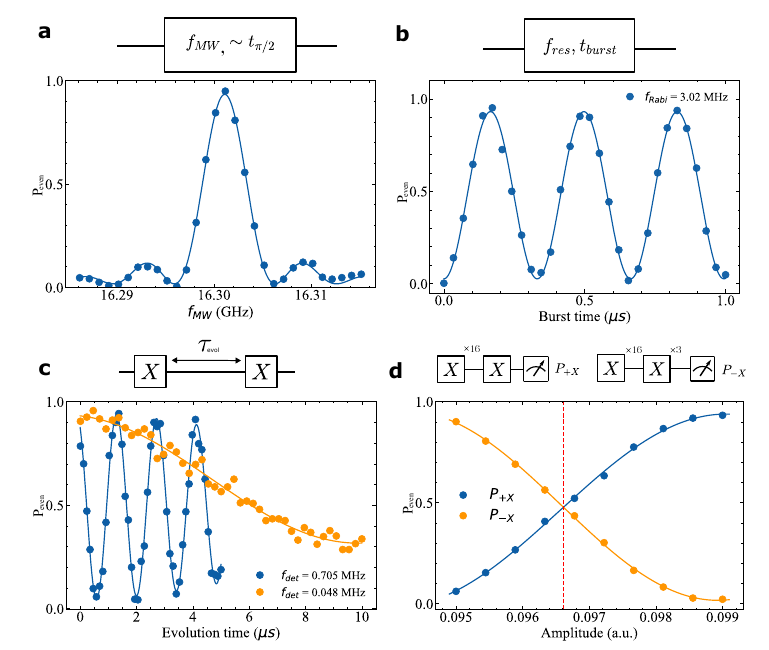}
\caption{\textbf{Single-qubit gate calibration.} 
\textbf{a} The resonance frequency is roughly tuned by applying a MW burst of gate time $t_{\text{g}}\sim t_{\pi}$ with varied frequency $f_{\text{MW}}$ and measure the returned even parity probability $\text{P}_{\text{even}}$. This probability is then fitted with equation $\text{P}_{\text{even}}(f_{\text{MW}}) = A \left( \frac{f_{\text{R}}^2}{(f_{\text{res}} - f_{\text{MW}})^2 + f_{\text{R}}^2} \sin ^{2} \left( \pi t_{\text{pulse}} \sqrt{(f_{\text{res}} - f_{\text{MW}})^2 + f_{\text{R}}^2} \right) \right) + B$ to obtain resonance frequency $f_{\text{res}}$. 
\textbf{b} The pulse amplitude is roughly tuned by measuring the Rabi oscillation with an amplitude $A$ and fit with $\text{P}_{\text{even}}(t_{\text{burst}}) = A\cos(2\pi f_{\text{R}}t_{\text{burst}})+C$ to get the Rabi frequency $f_{\text{R}}$. The pulse amplitude is adjusted to $A^{'}=A\times(f_{\text{R}}^{\text{target}}/f_{\text{R}})$ for a chosen target Rabi frequency $f_{\text{R}}^{\text{target}}$. The $\pi/2$ pulse gate time is set to $t_{\text{g}}=0.25/f_{\text{R}}^{\text{target}}$. 
\textbf{c} The frequency of the pulse is fine-tuned with a Ramsey sequence. We intentionally offset the MW frequency with a detuning of $\Delta f= 0.7$ MHz and apply the Ramsey sequence with pulse time $t_{\text{g}}$. The even parity probability $\text{P}_{\text{even}}$ of the Ramsey measurement oscillates with a frequency equal to the detuning from the resonance. We fit this oscillation with $\text{P}_{\text{even}}(t_{\text{evol}}) = A\cos(2\pi f_{\text{det}} t_{\text{evol}} +\phi) \exp(-(t/T_{2}^{*})^{2})+C$ and obtain the detuning $f_{\text{det}}$. We then shift the MW frequency towards the qubit resonance frequency by the amount of $\Delta f - f_{\text{det}}$. After this fine-tuning of the MW frequency, the Ramsey measurement shows a reduced detuning of $f_{\text{det}}=48$ kHz.
\textbf{d} To fine-tune the pulse amplitude, we apply a calibration sequence of 16 $\pi/2$ pulses, followed by either one or three additional $\pi/2$ pulses to rotate the spin into the $|+x\rangle$ or $|-x\rangle$ state, respectively. We then vary the pulse amplitude $V_{\text{amp}}$ and fit the resulting return probabilities with a sine function $\text{P}_{\text{even}}(V_{\text{amp}})=A\cos(kV_{\text{amp}}+\phi)+B$. The optimal amplitude is determined from the intersection point of the two fitted curves.
}\label{extend:gate_calib}
\end{figure}

\begin{figure}[H]
\centering
\captionsetup{font=footnotesize,skip=0pt,width=1\linewidth}
\includegraphics[width=1\textwidth]{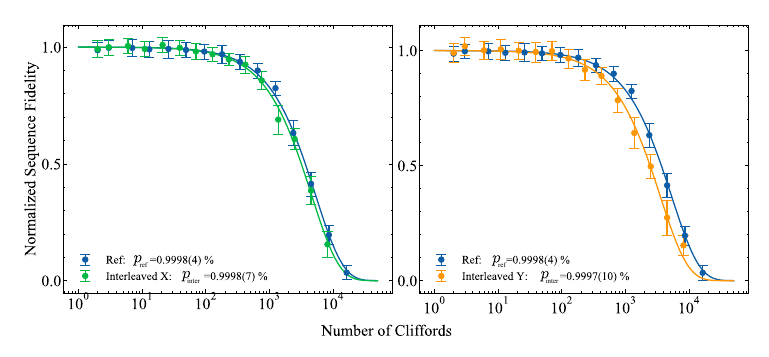}
\caption{\textbf{Single-qubit interleaved RB.} 
Measurements of the individual fidelity of the primitive $X(\pi/2)$ and $Y(\pi/2)$ single-qubit gates using interleaved randomized benchmarking. Because the $X$ and $Y$ gates differ only by a phase shift implemented in the AWG sequences that generate the MW signals using IQ mixers (Methods), they are expected to exhibit identical performance. The interleaved gate fidelity is calculated from the fitted depolarizing parameters of the reference $p_{\text{ref}}$ and interleaved $p_{\text{inter}}$ randomized benchmarking sequences using $F_{\text{inter}} = \frac{(d-1)(1+p_{\text{inter}}/p_{\text{ref}})}{d}$. The fitted depolarizing parameters are $p_{\text{ref}}=0.9998(4)$, $p_{\text{inter}}^{\text{X}}=0.9998(7)$ and $p_{\text{inter}}^{\text{Y}}=0.9997(10)$. The resulting primitive gate fidelities are 99.9985(8)\% for the X gate and 99.995(1)\% for the Y gate. Error bars of the data point represent the 95\% confidence interval and are calculated from the standard deviation of the data. Errors in primitive fidelities are obtained from the fitting covariance matrix. The error in the interleaved gate fidelity is calculated using error propagation.
}\label{extend:inter_rb}
\end{figure}

\begin{figure}[H]
\centering
\captionsetup{font=footnotesize,skip=0pt,width=1\linewidth}
\includegraphics[width=1\textwidth]{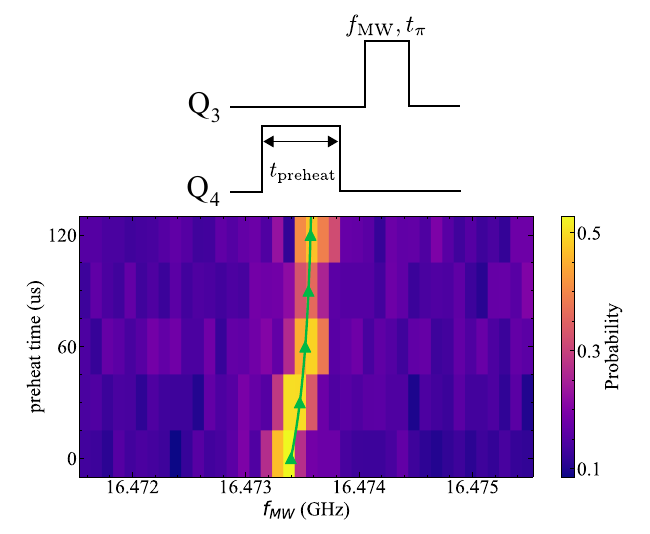}
\caption{\textbf{MW-Heating induced frequency shift.} 
Microwave-induced frequency shifts are commonly observed in Si/SiGe spin qubit experiments and can significantly impact the performance of quantum processors. Here, we show a resonance frequency shift of idling Qubit Q$_{3}$ when a pre-heating pulse is applied to Q$_{4}$ at a drive strength of $f_{\text{R}} = 3$~MHz. A frequency shift of approximately 0.22~MHz is observed for the longest heating pulse duration of $t_{\text{preheat}}=$120~$\mu$s. 
This measurement was performed at an elevated mixing chamber temperature of approximately 120 mK, consistent with the conditions used in the RB experiments presented in this work. At this temperature, MW-heating-induced frequency shifts have been reported to be reduced \cite{undseth_hotter_2023}. Green curve corresponds to an exponential model of $f_{0} + \Delta f_{\text{max}}(1-e^{-t_{\text{preheat}}/\tau})$ with $f_{0}=16473.4$ MHz, $\Delta f_{\text{max}}=200$ kHz and $\tau=60$ $\mu$s. We observe that these shifts scale with gate speed due to increased MW power, but remain significantly smaller than those reported in other experiments.}
\label{extend:heating_shift}
\end{figure}

\begin{figure}[H]
\centering
\captionsetup{font=footnotesize,skip=0pt,width=1\linewidth}
\includegraphics[width=1.\textwidth]{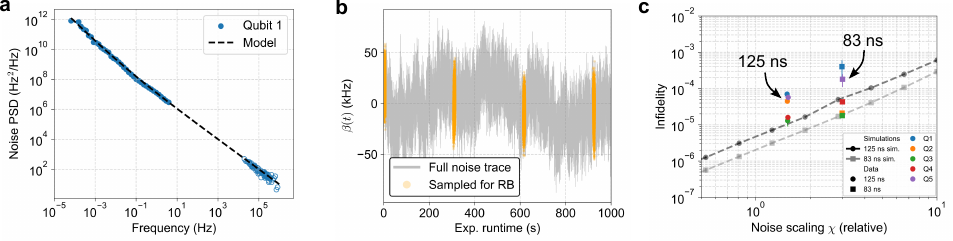}
\caption{\textbf{Simulations and noise-limited primitive gate fidelity.}
\textbf{a} Power spectral density (PSD) of qubit Q$1$, extracted from Ramsey (filled circles, low-frequency) and CPMG (open circles, high frequency) measurements \cite{yoneda_fast_2014, yoneda_noise-correlation_2023}, plotted as the positive-frequency part of the two-sided spectrum. Other qubits exhibit similar noise characteristics. A combined fit using $S(f) = 10^7(0.25\cdot f^{-1.3} + f^{-1})$ accurately reproduces the measured spectrum and is used to generate the simulated time-domain traces in (b). Notably, this PSD reflects noise under idling conditions, when the processor is not actively operated.
\textbf{b} Simulated low-frequency noise trace $\beta(t)$ generated from the PSD in (a), using the same parameters as in the experimental randomized benchmarking (RB) measurements: 500 single-shot measurements, 20 randomized sequences, 10 gate lengths, and a 1.75~ms RB cycle time for both recovery-gate sequences projecting into the up and down states. Long AWG loading times of approximately 5 minutes are also included. Orange traces indicate the subset of the noise used for RB simulations; only four repetitions are shown for clarity.
\textbf{c} Simulated primitive $\pi/2$ fidelities for different noise strengths. The noise scaling factor $\chi$ rescales the PSD as $S_{\chi}(f) = \chi^{2} S_{\chi=1}(f)$, corresponding to a linear scaling of the standard deviation. For gate times of 125 ns (83 ns), reproducing the experimental fidelities requires scaling the noise standard deviation by roughly 1.5× (3×). For longer gate times, the noise model captures the experimental fidelities rather well (see Fig.~\ref{fig2}(d)).
}
\label{extend:simulations}
\end{figure}

\begin{figure}[H]
\centering
\captionsetup{font=footnotesize,skip=0pt,width=1\linewidth}
\includegraphics[width=0.63\textwidth]{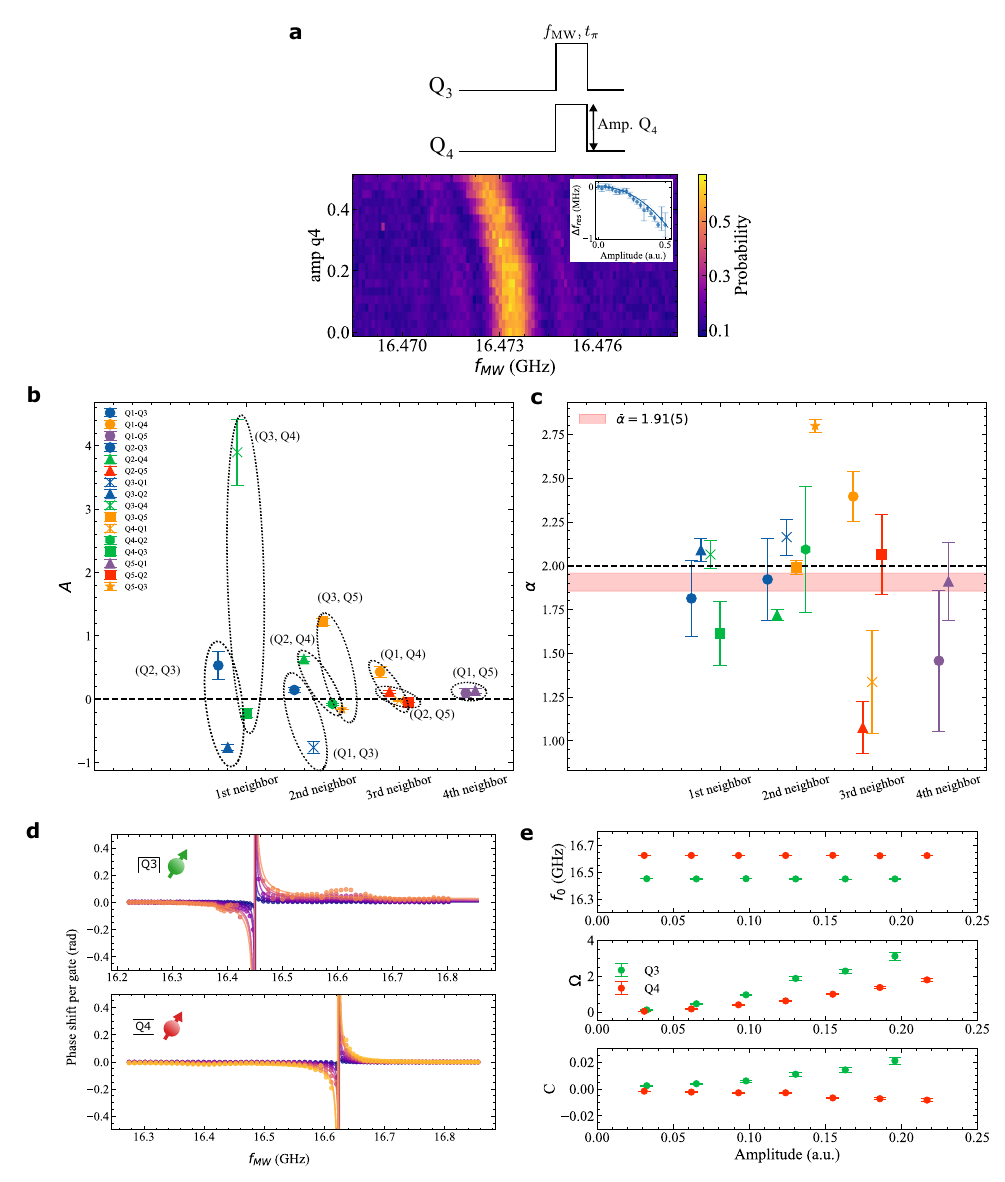}
\caption{\textbf{ {AC-Stark shift.}} 
\textbf{a} Resonance frequency shift of Qubit Q$_3$ when driven simultaneously with an $X^{2}$ gate on Q$_4$ (166~ns duration). The shift of $\sim$–0.8~MHz exceeds the ~$\sim$–0.2~MHz shifts caused by heating due to MW control pulses (see Extended Data Fig.\ref{extend:heating_shift}), indicating a distinct mechanism. Inset: Fit to the resonance frequency shift relative to $f_{\text{MW}}=16473.5$ MHz. The frequency shift is fitted with $A V_{\text{amp}}^{\alpha}$ resulting in $A=-3.2(3)$ and $\alpha=2.15(5)$.
\textbf{b, c}  Amplitude dependence of crosstalk-induced phase shifts between qubit pairs, fit with  $ A V_{\text{amp}}^{\alpha}$. The label Q$_{i}$-Q$_{j}$ indicates the phase shift on Q$_{i}$ when driving Q$_{j}$. The extracted average exponent $\alpha=1.91(5)$ is consistent with the expected quadratic scaling of the AC-Stark effect. As anticipated, the sign of the phase shift depends on detuning: it is positive when Q$_j$ is higher in frequency than Q$_i$ ($i < j$), and negative when lower ($i > j$), see panel (d).
\textbf{d} Crosstalk phase shift introduced by second-tone driving with fixed amplitude and varied frequency. As the second-tone frequency is swept across the qubit resonance, the induced phase shift changes sign. The observed phase shifts caused by the AC-Stark effect follow the form $\frac{\Omega^{2}}{x - f_{0}} + C$, where $\Omega$ is the driving strength, $f_{0}$ is the resonance frequency, and $C$ is a phenomenological offset — a positive (negative) detuning results in a positive (negative) phase shift.
\textbf{e} Fitted parameters from \textbf{d}. 
}\label{extend:crosstalk_phase}
\end{figure}

\begin{figure}[H]
\centering
\captionsetup{font=footnotesize,skip=0pt,width=1\linewidth}
\includegraphics[width=1\textwidth]{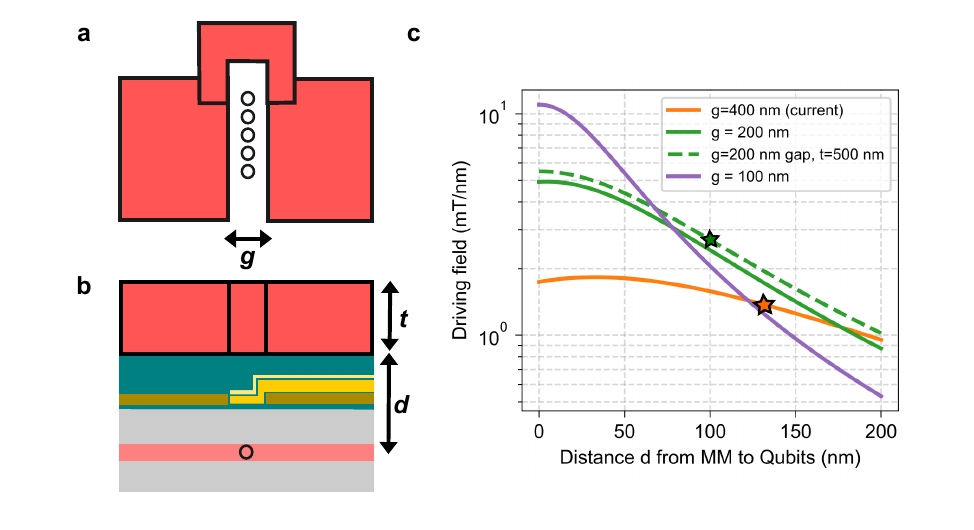}
\caption{
\textbf{Micromagnet design and simulated improvements in driving field gradient strength.}
\textbf{a} Top-down schematic of the micromagnet (MM) structure used for local spin control, with the micromagnet gap $g$ indicated \cite{yoneda_robust_2015}. Qubit positions are illustrated as circles. This illustration is not drawn to scale.
\textbf{b} Cross-section schematic of the device gate stack, showing three overlapping aluminum gate layers (shaded from brown to yellow) and the underlying 9~nm quantum well (red), where the qubits are located. The well is separated from the gates by a 30~nm SiGe spacer (gray). Oxides (pine-colored) electrically isolate the gate stack and the micromagnet. The MM thickness $t$ and qubit–MM separation $d$ are indicated for reference. This illustration is not to scale.
\textbf{c} Simulated slanting field gradients for different micromagnet gap sizes and thicknesses. The current device (400~nm gap, 250~nm thickness) achieves a slanting field of approximately 1.4~mT/nm at a vertical distance that spatially varies between 100–160~nm (orange star at 130~nm), depending on the number of underlying metal gate layers. For simplicity, we assume a flat surface in the simulations, corresponding to a uniform separation between the qubit and micromagnet. By reducing the qubit–MM separation to ~100~nm, through a thinner gate stack, and using a 200~nm gap, our simulations predict a potential doubling of the slanting field gradient. Further enhancement is possible by increasing the MM thickness, as demonstrated for the 200~nm gap and 500~nm thickness case. Stronger gradients enable faster gate operations at lower MW drive power.}
\label{extend:micromagnet_sim}
\end{figure}

\begin{table}[h]
\centering
\caption{\textbf{Overview of measured primitive gate fidelities.}
Primitive randomized benchmarking fidelities for all five qubits under various driving conditions, including isolated (1-qubit), pairwise (2-qubit), three-qubit (3-qubit), and fully parallel five-qubit operation (5-qubit). Control fidelities are shown for pulse durations of 83–500~ns used in these experiments.
}
\begin{tabular}{|c|c|c|c|c|c|}
\hline

Driving condition &  Q$_{1}$ & Q$_{2}$ & Q$_{3}$ & Q$_{4}$ & Q$_{5}$ \\ \hline

1-qubit (83 ns) & 99.96(2) & 99.9979(2) & 99.9982(2) & 99.9957(4) & 99.982(7)   \\  \hline
1-qubit (125 ns) & 99.993(1) & 99.9955(4) & 99.9987(4) & 99.9984(3) &  99.9943(7)  \\  \hline
1-qubit (250 ns) & 99.9979(1) & 99.9918(8) & 99.9977(8) & 99.9985(3) & 99.9943(7)  \\  \hline
1-qubit (500 ns) & 99.987(2) & 99.984(1) & 99.97(2) & 99.994(1) & 99.96(2)  \\  \hline
2-qubit (Q$_{2,3}$, 125 ns) & N/A & 99.9916(4) & 99.9930(2) & N/A & N/A  \\  \hline
2-qubit (Q$_{3,4}$, 125 ns) & N/A & N/A & 99.991(3) & 99.9970(4) & N/A  \\  \hline
2-qubit (Q$_{1,3}$, 250 ns) & 99.9937(6) & N/A & 99.991(1) & N/A & N/A  \\  \hline
2-qubit (Q$_{3,5}$, 250 ns) & N/A & N/A & 99.9950(11) & N/A & 99.9911(9) \\  \hline
3-qubit (Q$_{2,3,4}$, 250 ns)  & N/A & 99.9926(7) & 99.9929(16) & 99.9966(7) & N/A   \\ \hline
3-qubit (Q$_{1,3,5}$, 250 ns) & 99.992(1) & N/A & 99.995(2) & N/A & 99.983(2)  \\ \hline
5-qubit (250 ns) & 99.96(1) & 99.92(3) & 99.977(8) & 99.93(4) & 99.82(7)    \\  \hline
5-qubit (500 ns) & 99.978(4) & 99.91(2) & 99.96(2) & 99.94(5) & 99.8(1)    \\  \hline

\end{tabular}

\label{extend:all_fid}
\end{table}

\end{document}